\documentclass[letterpaper,aps,twocolumn,floatfix,amsfonts,amssymb,notitlepage,superscriptaddress,longbibliography]{revtex4-2}

\pdfpageattr {/Group << /S /Transparency /I true /CS /DeviceRGB>>}

\usepackage{amsmath}
\usepackage{amsthm}
\usepackage{amssymb}
\usepackage{graphicx}
\usepackage{tabularx}
\usepackage{color}
\usepackage{xcolor}
\usepackage[normalem]{ulem}
\usepackage[colorlinks=true, linkcolor=blue, citecolor=blue, urlcolor=blue]{hyperref}

\newcommand{\be}{\begin{equation}}
\newcommand{\ee}{\end{equation}}
\newcommand{\bea}{\begin{eqnarray}}
\newcommand{\eea}{\end{eqnarray}}

\def\bs#1\es{\begin{split}#1\end{split}}	\def\bal#1\eal{\begin{align}#1\end{align}}

\begin{document}

\title{Entropy of Non-Abelian Anyons from Slow Quasiparticle Dynamics in Quantum Hall Interferometers}

\author{Eran Sela}
\email{Corresponding author: eranst@tauex.tau.ac.il}
\affiliation{Raymond and Beverly Sackler School of Physics and Astronomy, Tel Aviv University, Tel Aviv 69978, Israel}

\author{Mitali Banerjee}
\affiliation{Institute of Physics, École Polytechnique Fédérale de Lausanne (EPFL), CH-1015 Lausanne, Switzerland}
\affiliation{Center for Quantum Science and Engineering (QSE Center), École Polytechnique Fédérale de Lausanne (EPFL), CH-1015 Lausanne, Switzerland}


\begin{abstract}
\noindent 
Non-Abelian anyons emerging in fractional quantum Hall states carry a characteristic entropy, $\Delta S = k_B \log d$, where \(d\) is the anyon's quantum dimension. This \(\mathcal{O}(1)\) entropy can, in principle, be extracted from charge measurements of an antidot via Maxwell relations. However, equilibrium charge measurements in fractional antidots have proven to be challenging with conventional charge detectors. 
Here, we propose a scheme based on an antidot embedded in an interferometer, in which the charge can be inferred from the recently observed time-dependent switching of the interference phase.
Performing such non-local charge measurements at equilibrium, the characteristic \(\mathcal{O}(1)\) entropy of non-Abelian anyons (e.g., $d = \sqrt{2}$ for the $\nu = 5/2$ state) can be extracted for intermediate temperatures, which exceed the
level spacing of the interferometer edge, but are much smaller than the level spacing of the antidot.
\end{abstract}

\maketitle


Moore and Read's prediction on the existence of non-Abelian anyons~\cite{moore1991nonabelions} marked a milestone in condensed-matter physics, particularly due to their promises in topological quantum computing~\cite{nayak2008non}. It raised the ubiquitous quest for experimental detection of nonabelian topological order. Beyond their non-commuting exchange statistics, it was soon realized that such gapped topological phases exhibit a ground-state degeneracy~\cite{wen1990ground}, which scales with the number of non-Abelian anyons as $g(N_{qp}) \sim d^{N_{qp}}$~\cite{Read_2000}. This implies an entropy of $S = k_B \log d$ per anyon, where $d>1$ is the quantum dimension of the non-Abelian anyon. While the fractional quantum Hall (FQH) state at filling factor $\nu = 5/2$ with charge $e^*=e/4$ nonabelian quasiparticles in GaAs long served as the primary candidate for a non-Abelian state~\cite{PhysRevLett.59.1776,Dolev_2008}, more recent experiments revealed numerous even-denominator states in graphene-based materials~\cite{Ki_2014,kim2015fractional,Zibrov_2017,Li_2017,Kim_2018,Kumar_2025} where such anyons are expected to appear. 

Several theoretical proposals have been concerned with the question of how to measure the entropy $S=k_B \log d$ associated with a single non-Abelian anyon~\cite{Yang_2009,cooper2009observable,gervais2010adiabatic,hou2012ettingshausen,Viola_2012,Schmidt_2017,ben2013detecting,schiller2025internal}.
In particular, let us emphasize the charging spectroscopy proposal of Ben-Shach \textit{et al.}~\cite{ben2013detecting}, based on earlier measurements of fractional charge at disorder-induced wells~\cite{Venkatachalam_2011}, which suggested to probe the entropy of localized non-Abelian quasiparticles in the $\nu = 5/2$ FQH state by exploiting the effect of temperature on the charge stability diagram. More recently, direct measurements of $\mathcal{O}(1)$ entropy in quantum dots were achieved by employing Maxwell relations and charge detection. This line of experiments started with the measurement~\cite{hartman2018direct} of the $k_B \log 2$ entropy associated with an electron spin in a quantum dot, and was generalized to strongly coupled quantum dots~\cite{PhysRevLett.129.227702}, bilayer graphene quantum dots with unique degeneracies~\cite{adam2024entropy},  double dots in GaAs/AlGaAs heterostructures~\cite{kealhofer2025entropydoublequantumdot,sheekey2026remoteentropymeasurementcoupled}, and exotic Kondo states~\cite{Piquard_2023,piquard2026experimentalevidencefractionalentropy}. The generality of this framework positions entropy as a promising spectroscopy method for several systems~\cite{han2022fractional,Ma_2023,PhysRevLett.131.016601,hurvitz2025metallicislandarraysynthetic}. Nevertheless, utilizing this general approach to measure the entropy $S=k_B \log d$ of non-Abelian anyons remains challenging.

In this work, we begin by presenting an ideal protocol to measure the entropy of anyons based on Ben-Shach's charging spectroscopy proposal~\cite{ben2013detecting}, or equivalently, the Maxwell relation approach in quantum dots. 
Extending quantum dots to the FQH regime requires a platform capable of controllably transporting individual fractional excitations into a dot structure. This can be achieved using \emph{antidots}, i.e. localized regions embedded within the FQH liquid which can therefore host nontrivial excitations~\cite{goldman1995resonant,Franklin_1996,Kataoka_1999,Kou_2012,Mills_2020,diluca2025quantumhallantidotfractional}. In the first part of this work we show that if the antidot's charge \(N(n_g, T)\) is directly controlled by a local gate voltage \(n_g\) and can be measured using a charge detector at thermal equilibrium, then 
one can directly infer the entropy of $k_B \log d$ 
in an appropriate regime.  

\begin{figure}
\centering
\includegraphics[width=\columnwidth]{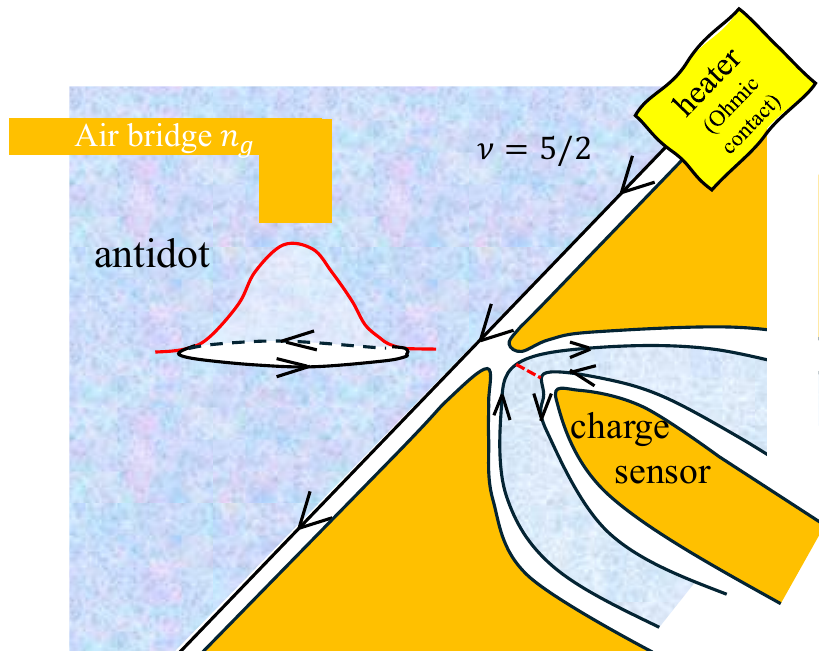}
\caption{Schematic of a device for measurement of non-Abelian entropy containing an antidot whose charge is controlled by a gate voltage $n_g$ and measured using a nearby charge sensor.
}
\label{fig:schematic0}
\end{figure}

Theoretically, it 
does not matter what method one uses to measure the average charge, and in this first part of the work, we have in mind a standard charge sensor, which is a conductor 
electrostatically coupled to 
the  antidot, see Fig.~\ref{fig:schematic0}. In this model, fractional charges tunnel to the antidot from a nearby long edge. By employing conformal field theory for both the long edge and the edge of the antidot, we show that for an intermediate temperature regime, much larger than the long edge's level spacings but much smaller than the energy level separation in the antidot, the entropy $\Delta S = k_B \log d$ can be directly inferred from the charge curve. While the $\nu = 1/3$ Laughlin state has only Abelian anyons giving $\Delta S = 0$, we discuss explicitly the non-Abelian $\nu = 5/2$ Pfaffian state where the anyon's quantum dimension $d_\sigma=\sqrt{2}$ determines the long sought fractional entropy.

However, charge measurements in fractional antidots using conventional electrostatically coupled charge detectors have proven to be challenging. The difficulty arises from the need for weak coupling between the antidot and the surrounding edge states to maintain a well-defined fractional charge; this weak coupling, in turn, leads to extremely long anyon tunneling times that were believed to exceed the stability time of the device. However, recently very long tunneling times have been measured in FQH interferometers, where the phase of the interference pattern exhibits slow dynamics both at $\nu=1/3$~\cite{Werkmeister_2025,samuelson2025slowquasiparticledynamicsanyonic}
and more recently in odd denominator states~\cite{kim2026selectivebraidingdifferentanyons}. These experiments are consistent with an antidot forming inside the interferometer loop whose fractional charge fluctuates slowly with time.

\begin{figure*}[t]
\centering
\includegraphics[width=\textwidth]{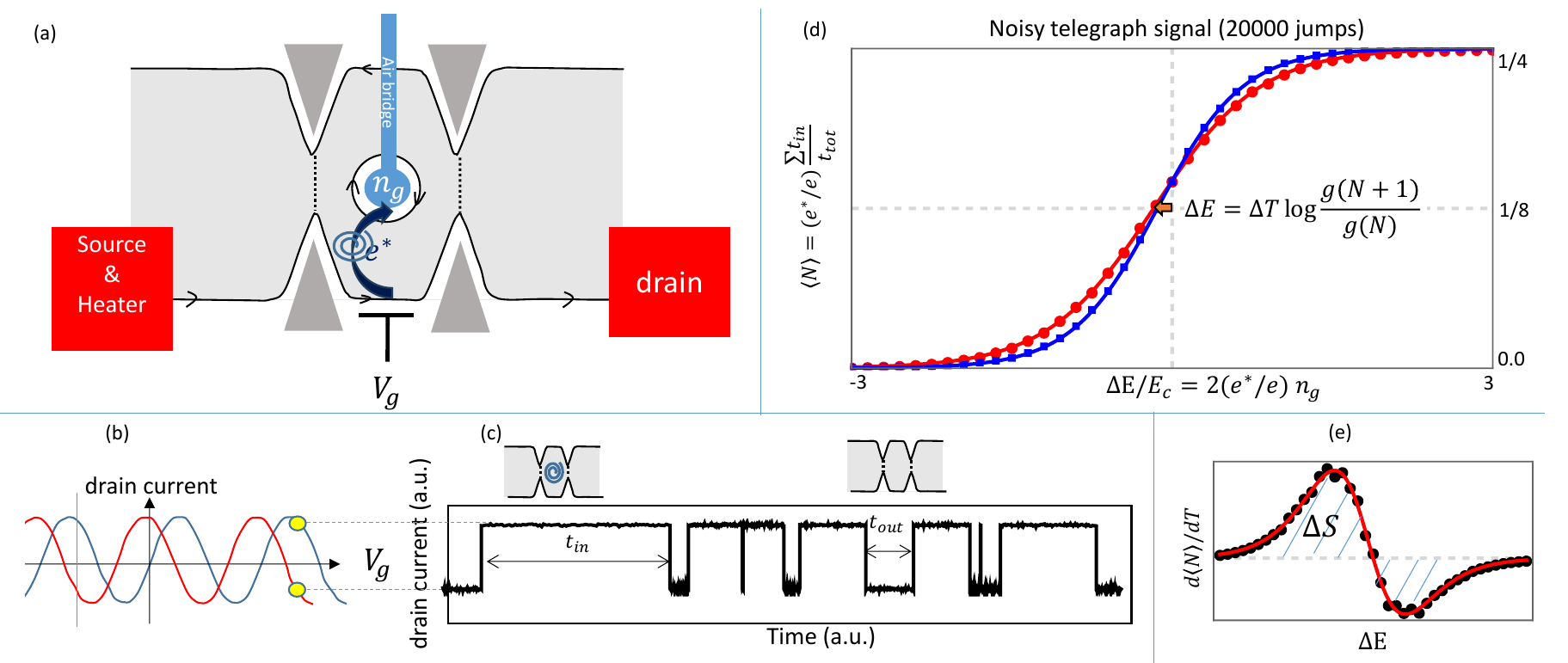}
\caption{Charge and entropy detection scheme via an antidot residing inside an interferometer. (a) The device contains a gate voltage $n_g$ directly controlling the number of quasiparticles in the antidot. (b) As a quasiparticle (charge $e^*$) tunnels into the antidot, the interference pattern as measured in the drain's current versus $V_g$ undergoes a phase shift. (c) Telegraph signal in the drain current, obtained by parking all device parameters to constant values  [calculated for $T/E_c=0.4$ and $\Delta E /E_c=0.25$]. It allows for extracting the relative times the QP spends inside ($t_{in}$) or outside ($t_{out}$) the antidot. (d) Illustration of charging curve extracted from telegraph signal, obtained by plotting $\frac{\sum_i t_{in}}{\sum_i t_{in}+\sum_i t_{out}}$  for many values of $n_g$, at $T/E_c=0.4,0.5$. When varying the temperature (using the heater in (a)), the curve broadens as well as shifts by an amount directly related to the anyon's entropy $g(N)$. (e) Upon integration, the difference of charge curves at temperatures $T$ and $T+\Delta T$ gives the entropy difference in the presence and absence of an anyon in the antidot.
}
\label{fig:schematic}
\end{figure*}

Here, we propose to exploit this phase switching as a non-local probe in a controlled and gated antidot device, from which the equilibrium charge and hence the entropy can be extracted. More broadly, the Aharonov-Bohm interference serves as a \emph{non-local charge detector}, which removes the need for a conventional nearby charge detector with high sensitivity. As we describe, by monitoring the temperature dependence of the interference switching, one can effectively extract the charge curves as in the original Ben-Shach's charging spectroscopy proposal, see Fig.~\ref{fig:schematic}. From it, the entropy directly follows.

\textbf{Setup and Maxwell relation:} 
We consider an antidot where the localized quantum Hall edge states are electrostatically coupled to a gate parameter $n_{g}$. We apply a standard effective Hamiltonian
\be
H_c=E_c(N-n_g)^2,
\ee
where $E_{c}$ represents the effective addition energy coefficient (or inverse compressibility) of the antidot, comprising both local electrostatics and edge reconstruction configurations governing the addition of localized charge, rather than a rigid geometric capacitance.
As a function of time, the charge 
$N(t)$ (in units of $e$)
can fluctuate by discrete fractional values as tunneling of quasiparticles from a nearby edge takes place. Its thermal average  $\langle N \rangle_{T,n_g}$ is then measured over long times, describing a thermal equilibrium state, as a function of temperature $T$ and gate voltage $n_g$. A device realizing this setup is shown in Fig.~\ref{fig:schematic0}, where the antidot is gated via an air bridge carrying the gate voltage \(n_g\).

At filling factor \(\nu = 5/2\) (and related even-denominator states), the antidot admits charge states quantized in units of \(e^*=e/4\). 
In this section, we address the question: what is the total entropy change associated with the gradual insertion of a  \(e/4\) quasiparticle into the antidot?  Namely, we consider a continuous change of $n_g$ such that $\langle N \rangle_{T,n_g}$ goes from one charge plateau to another one, differing by $e/4$, and ask what the total entropy difference is.

Let us start with a pictorial description of the thermodynamics of this system, treating the tunneling of quasiparticles as a classical stochastic process (see Fig.~\ref{fig:schematic}). The partition function of the antidot is $Z=\sum_{N} g(N) e^{-E_c(N-n_g)^2 /T}$. Here $N$ consists of all possible fractional charges of the antidot allowed by the FQH state, namely $N=\frac{e^*}{e}N_{qp}$ where $N_{qp}$ is the number of quasiparticles, and $g(N)$ is the degeneracy of the corresponding charge state. Using this partition function near a charge degeneracy between a state with $N_{qp}$ or $N_{qp}+1$ quasiparticles, for $E_c \gg k_B T$, it is sufficient to retain only two charge states $N=N_0$ and $N=N_0+e^*/e$. Then, within this  classical stochastic description, the average charge  behaves like a Fermi function up to a scale factor of $e^*/e$,
\be
\label{eq:charge_curve}
\langle N \rangle_{T,n_g} \equiv N_0+\frac{e^*}{e} \frac{1}{1+e^{-\Delta E(n_g)/k_B T}\frac{g(N)}{g(N+e^*/e)}},
\ee
where $\Delta E=H_c(N_0) - H_c(N_0+e^*/e)=2 (e^*/e)E_c (n_g-e^*/2e)$. As illustrated in Fig.~\ref{fig:schematic}(d) below, for any fixed temperature, this is nothing but a rescaled and shifted Fermi function.  
The extra factor $\frac{g(N+e^*/e)}{g(N)}$ is the ratio of degeneracies as one extra quasiparticle is added. Without this extra degeneracy factor, the middle of the charge step occurs at $\Delta E=0$. But when  $\frac{g(N+e^*/e)}{g(N)}>1$, upon increasing the temperature by $\Delta T$, the charge curve shifts to the left (negative $n_g$) by 
\be
\Delta E =-k_B \Delta T  \log \frac{g(N+e^*/e)}{g(N)},
\ee
see Fig.~\ref{fig:schematic}(d).
Thus, from the shift of the measured charge curve in gate voltage with $\Delta T$, one can extract the degeneracy factor~\cite{ben2013detecting}. The more general statement follows from the Maxwell relation: the entropy difference as the gate voltage is swept from $n_{g1}$ to $n_{g2}$ is
\be
\label{eq:Maxwell}
\Delta S =k_B E_c \int_{n_{g1}}^{n_{g2}} dn_g \frac{d\langle N \rangle}{dT}.
\ee
As we go from one charge plateau to the other, the entropy difference coincides with $k_B  \log \frac{g(N+e^*/e)}{g(N)}$.

In the above analysis, the nontrivial properties of the FQH state are encoded in the degeneracy factors $g(N)$, and otherwise the same analysis is applicable for electrons. For bare electrons with spin, this shift of the charging curve allowed Hartman \textit{et al.}~\cite{hartman2018direct} to measure the $k_B \log 2$ spin degeneracy.

\textbf{Isolating the entropy of a single  anyon:}  While the above argument is essentially the one given by Ben-Shach et. al.~\cite{ben2013detecting}, it sweeps under the rug a fundamental question:
When a fractional quasiparticle is added to the antidot, a compensating fractional charge must be created elsewhere.
But then, the degeneracy factor $g(N)$ should be associated with both of these fractional excitations. This is expected to spoil the fractional nature of the entropy change. 
Under what conditions, if any, does the total entropy difference of this process of charging the antidot encode the entropy $k_B \log d$ of a single anyon? 

The key idea to discriminate the entropy contribution of the nonabelian anyon added to the antidot, and the corresponding anyon which must be created elsewhere, is to introduce an energy scale that will separate them. A similar challenge was addressed in a recent proposal to measure the entropy of a single Majorana fermion in the Kitaev chain, which always hosts a pair of end modes~\cite{sela2019detecting}. There, it was suggested to strongly hybridize one of the two end Majoranas with a lead, thereby quenching its entropy for temperatures smaller than the hybridization. Here, we create an energy separation between the two anyons by assuming a large level spacing in the antidot as compared to the long edge. Based on a conformal field theory analysis of the edge partition function, we show that for temperatures $T$ which exceed the small level spacing of the long edge, but are much smaller than the level spacing of the antidot, the total entropy difference between $N_0$ and $N_0+e^*/e$ is exactly given by 
\be
\label{eq:Delta_S_charge_spectrostocpy}
\Delta S=k_B  \log \frac{g(N_0+e^*/e)}{g(N_0)} = k_B \log d,
\ee
where $d$ is the quantum dimension of the anyon added to the antidot.
The detailed derivation of this result is relegated to the End Matter.  

\begin{figure}
\centering
\includegraphics[width=0.9\columnwidth]{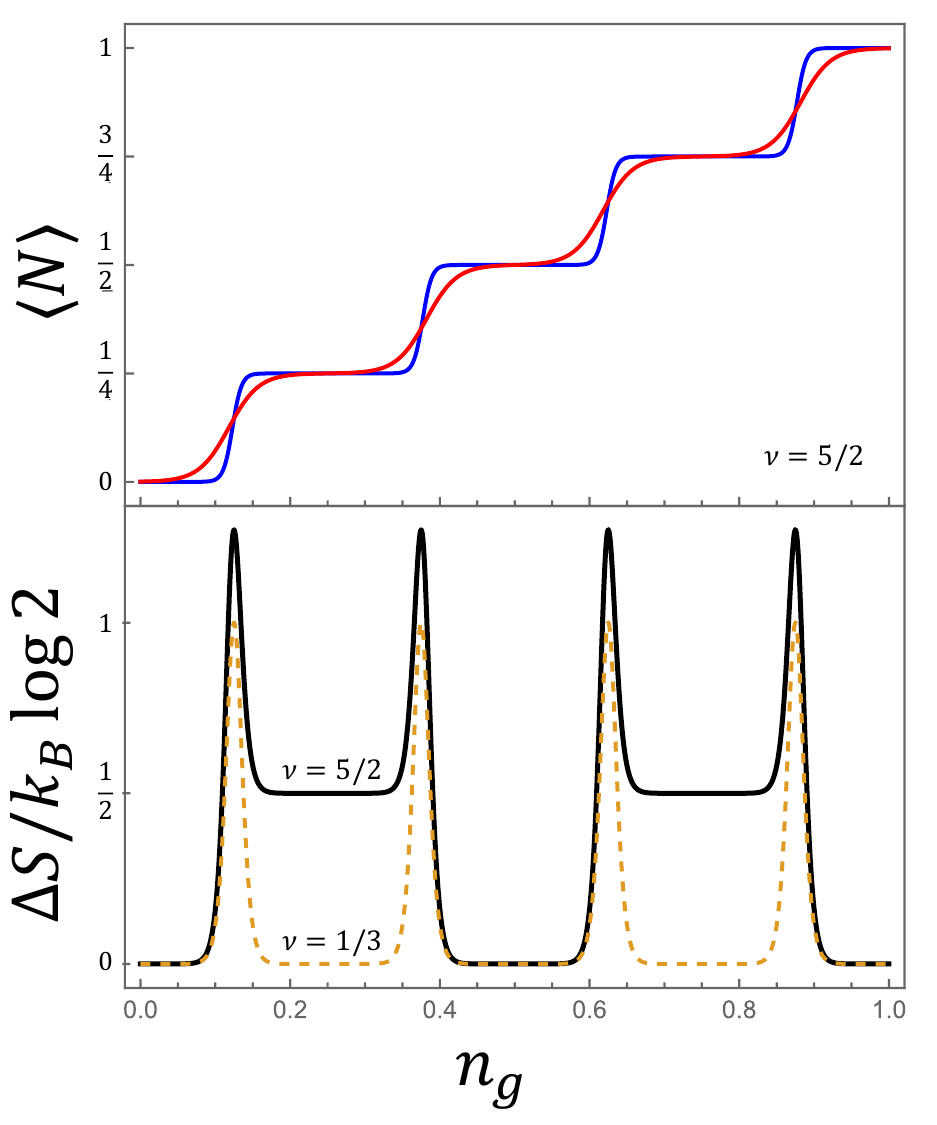}
\caption{$\frac{e}{4}$ charge steps, and resulting entropy obtained by integrating the Maxwell relation for $\nu=5/2$. The entropy displays peaks describing charge transitions. Between the charge transitions, we obtain an entropy of $\frac{1}{2}k_B\log 2$ when the antidot's charge is an odd multiple of $e/4$, namely when it contains unpaired nonabelian anyons. In contrast, for $\nu=1/3$ (or any abelian state), we obtain just a sequence of $k_B \log 2$ peaks. 
}
\label{fig:Fig_small_antidot_entropy}
\end{figure}

Based on this result, the charge curves given by Eq.~(\ref{eq:charge_curve}) generalized to few charge steps, and the entropy change obtained by integrating $\frac{d \langle N \rangle}{dT}$ using the Maxwell relation Eq.~(\ref{eq:Maxwell}), are shown in Fig.~\ref{fig:Fig_small_antidot_entropy}. For the $\nu=5/2$ state with charge $e^*/e=1/4$ anyons, carrying the quantum dimension $d_\sigma=\sqrt{2}$, we obtain an even-odd effect: the states with an unpaired anyon in the antidot have an extra entropy of $\frac{1}{2}k_B \log 2$. We note that this extra entropy is not associated with that of the antidot, due to the large level spacing in the antidot, but rather with the remote anyons that must be created in the long edge due to the fractional occupation of the antidot. Right at the charge degeneracy, there is an extra entropy peak of $k_B \log (1+\sqrt{2})$ separating the two different entropy values of the paired and unpaired states. On the other hand, switching to $\nu=1/3$, all the anyons are abelian, $d=1$, and there is no change in degeneracy factor.

\textbf{Interference as a fractional charge sensor}: 
So far, we have assumed that it is possible to measure the charging curve of a fractional antidot. However, this turns out to be challenging using conventional charge sensors. 

Here, we rely on the key observation that the charge of the antidot can influence the interference current via the non-local Aharonov-Bohm effect. This was recently observed in a Mach-Zehnder type interferometer using a gate-controlled antidot~\cite{ghosh2025anyonic,ghosh2025coherentbunchinganyonsdissociation}. In these experiments, the current was not measured as a function of time. When the gate voltage of the antidot is close to a charge degeneracy, one expects a telegraphic signal, exactly as previously measured in electron tunneling events in quantum dots~\cite{Gustavsson_2006} and, as discussed above, now also observed in FQH interferometers~\cite{Werkmeister_2025,samuelson2025slowquasiparticledynamicsanyonic,kim2026selectivebraidingdifferentanyons}.

From the observed time-dependent telegraphic signal associated with anyon tunneling, one can measure the charge curves versus gate voltage and temperature. As we schematically illustrate in Fig.~\ref{fig:schematic} based on recent experiment~\cite{werkmeister2025anyon}, the interference current can act as a charge detector of fractional charges - it can switch in a telegraph noise-like fashion between few specific values as quasiparticles tunnel into or out of the antidot, very much like electrostatics based charge sensing of electrons tunneling into an out of a quantum dot~\cite{Gustavsson_2006}. Thus, we envision the possibility to use interference based charge detection [Fig.~\ref{fig:schematic}(d)]. In panel (c), we simulate a classical stochastic switching between two states corresponding to an empty state $(0)$ and a state occupied by one anyon $(1)$. We assume an equilibrium Markov process with detailed balance such that $\frac{\Gamma_{0 \to 1}}{\Gamma_{1 \to 0}}=e^{\Delta E/T}\frac{g(N_0+e^*/e)}{g(N_0)}$. We disregard any coherent superposition of these two states~\cite{mross2026anyonzenoeffect} by assuming that the tunneling energy scale, corresponding to very large tunneling times, is much smaller than the thermal scale. The resulting telegraphic signal consists of a collection of time segments of duration $t_{in}$ where the quasiparticle occupies the antidot, as measured by a particular value of the interference phase in (b), followed by jumps to an unoccupied state of duration $t_{out}$.  To extract the charge curve $\langle N \rangle$, we mimic the experimental protocol and repeat this collection of telegraph signals for a different temperature, as obtained by heating the edge. From our knowledge of the charge of fractional quasiparticles $e^*$ we deduce the charge of the antidot to be $N=N_0+(e^*/e) \frac{\sum_i t_{in}}{\sum_i t_{in}+\sum_i t_{out}}$. From this charge curve, as we showed above, the entropy of the quasiparticles and particularly their quantum dimension can be extracted using the Maxwell relation, see Fig.~\ref{fig:schematic}(e).

\textit{Conclusion.--}  
Detecting non-Abelian topological order experimentally remains one of the biggest open challenges. Among the most direct experiments advances to date is the measurement of the fractionalized quantum heat conductance in the $\nu=5/2$ non-Abelian state~\cite{Banerjee_2017}, which reflects the {\emph{entropy-density}} of the edge. In this work, we proposed a measurement of the entropy of order $k_B$ of individual nonabelian quasiparticles, via tunneling between an edge and an antidot, accompanied by charge detection and controlled temperature variations. Our theoretical advance is the establishment that from charge measurements, one can extract the entropy of a single nonabelian anyon. As a further experimental perspective, our work opens the way towards charge measurements of the equilibrium thermodynamics of nonabelian anyons, using time-dependent switches in interference.


\begin{acknowledgments}
\noindent{\textit{Acknowledgments.--}} We thank inspiring discussions with Klaus Ensslin and Joshua Folk, which triggered the quest for alternative charge sensors of antidots. We also thank Ady Stern and, particularly, Moty Heiblum for suggesting the experimental possibility to detect charges via slow time-dependent interference. MB acknowledges discussion with Emily Hajigeorgiou regarding the experimental setup. ES gratefully acknowledges support from the European Research Council (ERC) under the European Union Horizon 2020 research and innovation programme under grant agreement No. 951541. M.B. acknowledges the support of the SNSF Eccellenza grant No. PCEGP2\_194528.
\end{acknowledgments}


\section{END MATTER}
\textbf{Thermodynamics of a single edge:} 
Let us recall basic thermodynamic properties of an edge of a FQH liquid. Let us consider a chiral edge of length $L$ with periodic boundary conditions. $L$ is assumed to be larger than any microscopic length scale such that the continuum conformal field theory (CFT) description is valid. The level spacing is set by \(2\pi/L\) (with \(\hbar=1\) and with the edge velocity set to unity). We first focus on a Laughlin state at filling factor \(\nu = 1/3\). Then the edge admits three topological sectors labeled by
\(a = 0, \pm 1/3\), corresponding to the fractional part of the edge charge.
We define \(q = e^{- 2\pi\beta/L}\) such that the latter condition implies \(q \to 1\). Then the partition function in sector \(a\) is described by a chiral free-boson CFT (see e.g. Ref.~\cite{fendley2007topological}),
\be
\label{eq:chi_freeboson}
\chi_a(q) =\frac{1}{\eta(q)} 
\sum_{n\in\mathbb{Z}} q^{\frac{3}{2}(n+a)^2},
\ee
where $\eta(q)=q^{c/24} \prod_{m=1}^\infty (1 - q^m)$ is the Dedekind function, and the central charge $c=1$ for the Laughlin state.
The sectors \(a=\pm 1/3\) are accessed by creating a compensating fractional charge \(\mp 1/3\) in the bulk, while the ground-state edge partition function is \(Z_{\mathrm{edge}}=\chi_0(q)\).

The temperature $k_B T=\beta^{-1}$ is assumed to be much smaller than the bulk gap. But it can be larger or smaller than the level spacing of the edge. Here, we are interested in the regime $k_B T \gg \frac{2\pi}{L}$. Then $q \approx 1$, and many terms contribute to the sum in Eq.~(\ref{eq:chi_freeboson}). To evaluate the partition function in this regime, we use the fact that the CFT partition functions exhibit a modular symmetry between space and time, which exchanges the spatial length \(L\) and the inverse temperature \(\beta\).
Introducing \(\tilde{q} = e^{-2\pi L/\beta}\), the partition functions transform as $\chi_a(q) = \sum_b S_{ab}\,\chi_b(\tilde{q})$ 
where \(S_{ab}\) is the modular \(S\)-matrix.
In our regime of interest $k_B T \gg \frac{2\pi}{L}$ we have \(\tilde{q} \ll 1\). Then the term \(b=0\) corresponding to the identity sector dominates since it contains the smallest power of $\tilde{q}$. Therefore,
\be
\label{eq_chi_a}
\chi_a(q)  \xrightarrow{q \to 1} 
 S_{a 0} \tilde{q}^{-c/24} = S_{a 0} e^{\pi c L k_B T/ 12}.
 \ee
This result depends only on the central charge of the CFT and on the associated modular $S-$matrix and associated topological sectors.
This immediately gives the free energy (reinserting the velocity $v$)
\be
\label{eq:F}
F_a = -k_B T \log \chi_a(q)= -\frac{\pi c L}{12 v}(k_B T)^2 - k_B T \ln S_{a0},
\ee
and the corresponding entropy $S_a = \frac{\pi c L}{6 v} k_B^2 T+ k_B \log S_{a0}$. The last term is known as the Affleck-Ludwig boundary term. In the ground state sector $a=0$ we have $F_{\mathrm{edge}} = -k_B T \log Z_{\mathrm{edge}}$ giving $S_{\mathrm{edge}}=\frac{\pi c L}{6 v} k_B^2 T$.

One of the direct confirmations of the chiral CFT description is the experimental observation of quantized thermal transport. In particular, the heat current
\[
J_Q = \frac{\pi c}{6}(k_B T)^2,
\]
was first measured for an integer quantum Hall edge~\cite{jezouin2013quantum}. Interestingly, while a fractional Laughlin state has the same quantized heat conductance as the \(\nu=1\) state, the situation changes for non-Abelian edges. For the Moore--Read (MR) state, the chiral central charge is \(c = 3/2\), with the extra \(c = 1/2\) arising from the non-Abelian Ising sector. Other candidate states predict different total chiral central charges (\(c=-3/2\) for the anti-Pfaffian~\cite{Levin_2007} and \(c=1/2\) for the PH-Pfaffian~\cite{Son_2015}), but they all share the half-integer contribution from the Ising non-Abelian sector. This half-integer contribution has been observed experimentally via heat conductance measurements~\cite{Banerjee_2017}.

While the CFT description is strongly predictive for edge thermodynamics, these experiments are insensitive to the boundary term in the free energy involving the modular \(S\)-matrix, which encodes the quantum dimension of the underlying anyons, 
\be
\label{eq:Sa00d}
S_{a0}=S_{0a}=d_a/\mathcal{D},
\ee
where $\mathcal{D}=\sqrt{\sum_a d_a^2}$ is the total quantum dimension. In Eq.~(\ref{eq:F}) only the identity sector enters with $S_{00}=1/\mathcal{D}$. As we will see, a nontrivial contribution from the boundary term becomes directly observable when tunneling into an antidot is introduced.

\textbf{Two-edge CFT thermodynamics:} We model the antidot as a short edge of length \(L_{\rm AD}\), described by the CFT Hamiltonian \(H_{L_{\rm AD}}\), and include its charging energy \(H_c\), so that the total Hamiltonian is \(H_{{L_{\rm AD}}} + H_c\). Similar to $q$, we define
\(p = e^{-2\pi \beta / L_{\rm AD}}\). Since the antidot can be small, below we will consider $p \ll 1$.

To incorporate the antidot’s charging energy, we define \emph{charge-resolved characters}
\[
\chi_a(p, N) = \int_0^{2\pi} \frac{d\varphi}{2\pi} \, {\rm tr}_a \left[ e^{i \varphi (\hat{N}-N)} e^{-\beta L_{\rm AD}} \right].
\]  
For each fixed \(N\), \(\chi_a(p, N)\) corresponds to a particular term in the sum over \(n\) in Eq.~(\ref{eq:chi_freeboson}).

The total partition function of the system is
\be
\label{eq:general_Z}
Z(n_g) = \sum_N e^{-\beta E_c (N-n_g)^2} \sum_{a \in N} \chi_a(p, N) \chi_{\bar{a}}(q),
\ee
where the sum over \(a \in N\) runs over all antidot topological sectors compatible with total charge \(N\), and \(\bar{a}\) labels the conjugate sector of the long edge. This expression accounts for thermodynamic fluctuations of the antidot charge across all allowed values, including fractional charges determined by the underlying FQH state. 

In writing this partition function, we assume that tunneling between the antidot and the long edge is weak, so that no coherent superpositions of states with different \(N\) are formed. To proceed, let us consider as examples the states $\nu=1/3$ and $\nu=5/2$.

\textbf{Laughlin state $\nu=1/3$:} In this case the charge of the antidot, $N$, can be any multiple of $1/3$. In such an abelian state, we have $d_a=1$ for all anyon types.
Consider the charge transition: $N=0 \to 1/3$. When the charging energy is large, we keep only these two states in Eq.~(\ref{eq:general_Z}), 
\bea
Z_{0\to \frac{1}{3}}(n_g)&=&e^{-\beta E_c (0-n_g)^2} \chi_0(p,0)\chi_0(q) \nonumber \\
&+&e^{-\beta E_c (\frac{1}{3}-n_g)^2} \chi_{1/3}(p,{\scriptscriptstyle \frac{1}{3}})\chi_{-1/3}(q).
\eea
Let us use Eq.~(\ref{eq_chi_a}) to evaluate the $q$-dependent long edge partition functions, giving $\chi_a(q \to 1)=S_{a 0} e^{\pi c L k_B T /6}$. Since all the $d_a$'s are equal to each other, we see from Eq.~(\ref{eq:Sa00d}) that $\chi_{0}(q)$ and $\chi_{-1/3}(q)$ become equal in the long edge limit. Therefore, the long edge partition functions do not contribute to the charge profile $N(n_g,T)$ and can be factored out,
\bea
Z_{0\to \frac{1}{3}}(n_g)&\propto& e^{-\beta E_c (0-n_g)^2} \chi_0(p,0) +e^{-\beta E_c (\frac{1}{3}-n_g)^2} \chi_{1/3}(p,{\scriptscriptstyle \frac{1}{3}}). \nonumber
\eea
The charge resolved $p$-dependent antidot partition functions, using Eq.~(\ref{eq:chi_freeboson}), are given by
\be
\chi(p,N)=\frac{1}{\eta(p)}
\begin{cases}
1, & \text{if } N = 0,\\[2mm]
p^{1/6}, & \text{if } N = 1/3.
\end{cases}
\ee
The Dedekind function $\eta(q)$ also factors out. The piece $p^{1/6}=e^{-2\pi \beta/(6 L_{AD})}$ enters the partition function simply as a temperature-independent shift of the degeneracy point between charges $N=0$ and $N=1/3$. It becomes negligible when $E_c \gg \frac{2\pi}{L_{AD}}$ and we ignore it. Then the two terms in the resulting partition function are of the form of Eq.~(\ref{eq:Delta_S_charge_spectrostocpy}) with equal degeneracy factor for the two charge states. The resulting charge steps and associated entropy curve obtained by integrating the Maxwell relation are trivial, having just $k_B \log 2$ entropy peaks at the charge degeneracy points, see dashed curve in Fig.~\ref{fig:Fig_small_antidot_entropy} (bottom). 

\textbf{MR state:} Now consider the non-Abelian MR state with $c=3/2$. 
Using the notation of Ref.~\cite{fendley2007topological}, the theory is the tensor product of a free boson theory with operators $V_{\gamma}=e^{i \gamma \varphi(z)}$ having charge $\gamma/\sqrt{2}$, and an Ising theory with operators $I,\psi,\sigma$ being charge-neutral. The consistent combined theory has 6 primary fields
\be   
a \in \{ I, \sigma V_{\pm \frac{1}{2\sqrt{2}}},\psi,V_{\pm 1/\sqrt{2}} \},
\ee
corresponding to charge excitations $\{0,\pm { \frac{1}{4}} ,0, \pm { \frac{1}{2}}\}$. (Here we label the vacuum sector by $I$ rather than $0$). The corresponding anyon quantum dimensions are
\be
d_a=\{1, \sqrt{2}, \sqrt{2},1,1,1 \},
\ee
with $\sigma$ being the nonabelian anyon. The total quantum dimension then is $\mathcal{D}=2\sqrt{2}$.

Consider the partition function describing the charge transition $N=0 \to 1/4$. The $N=0$ charge state may be in the two topological sectors $I,\psi$, and the state $N=1/4$ must be in the topological sector $\sigma V_{+\frac{1}{2\sqrt{2}}}$. To find the corresponding topological sectors of the long edge, we use the fusion rules  $\psi \times \psi = I$, and $\sigma \times \sigma=I+\psi$, giving $\bar{\psi}=\psi$ and $\bar{\sigma}=\sigma$. Hence, the two-charge state reduced partition function is
\bea
Z_{0 \to {\scriptscriptstyle \frac{1}{4}}}(n_g)&=&e^{-\beta E_c (0-n_g)^2} [\chi_I(p,0)\chi_I(q)+\chi_\psi(p,0)\chi_\psi(q)]\nonumber \\
&+&
e^{-\beta E_c (\frac{1}{4}-n_g)^2} \chi_{\sigma V_{\frac{1}{2\sqrt{2}}}}(p,{\scriptscriptstyle \frac{1}{4}})\chi_{\sigma V_{- \frac{1}{2\sqrt{2}}}}(q).
\eea
In this case, using Eq.~(\ref{eq_chi_a}) to evaluate the $q$-dependent long edge partition functions, we observe that the factor $S_{aI}$, determined by the anyon's quantum dimensions, gives a larger weight to the contributions of the charge $1/4$ state which has $d_\sigma = \sqrt{2}$ as opposed to the zero charge state which contains only abelian anyons with $d_a=1$. Taking the common factor out, we have
\bea
Z_{0 \to {\scriptscriptstyle \frac{1}{4}}}(n_g)&\propto &e^{-\beta E_c (0-n_g)^2} [\chi_I(p,0)d_I+\chi_\psi(p,0)d_\psi]\nonumber \\
&+&
e^{-\beta E_c (\frac{1}{4}-n_g)^2} \chi_{\sigma V_{\frac{1}{2\sqrt{2}}}}(p,{\scriptscriptstyle \frac{1}{4}})d_\sigma.
\eea
The charge resolved $p$-dependent antidot partition functions for $N=0,1/4$ are given by
\be
\chi_{\sigma V_{\frac{1}{2\sqrt{2}}}}(p,{\scriptscriptstyle \frac{1}{4}})=\frac{\chi_\sigma(p)p^{1/16}}{\eta(p)},~~~\chi_{I,\psi}(p,0)=\frac{\chi_{I,\psi}(p)}{\eta(p)}.
\ee
As above, the Dedekind functions factor out, and the factor $p^{1/16}$ 
gives a temperature-independent shift of the degeneracy point. We are left with the extra partition functions $\chi_{I,\psi,\sigma}(p)$ from the Ising CFT.

We now consider a short antidot edge so that $p \ll 1$. Then these partition functions are dominated by their ground state, 
\be
\chi_{a}(p)  \xrightarrow{p \to 0} p^{h_a - \frac{c}{24}},
\ee
with $h_{a=I,\psi,\sigma}=(0,1/2,1/16)$. When the charge is $N=0$, there are two competing states $I,\psi$. But when $T \ll \frac{2\pi}{L_{AD}}$, only the ground state $I$ dominates. The partition function becomes  
\bea
Z_{0 \to {\scriptscriptstyle \frac{1}{4}}}(n_g)\propto e^{-\beta E_c (0-n_g)^2} d_I+
e^{-\beta E_c (\frac{1}{4}-n_g)^2} d_\sigma .
\eea
Now we can see that the charge state $N=1/4$ has an extra degeneracy given by $d_\sigma / d_I=\sqrt{2}$. As shown in Fig.~\ref{fig:Fig_small_antidot_entropy}, this results in an entropy change between the $N=0$ and $N=1/4$ charge plateaus of $\Delta S = \frac{1}{2} k_B \log 2$. At the peak, the entropy is $k_B  \log (d_I+d_\sigma)$.

Physically, this result uses the smallness of the dot. In this limit, due to a large level spacing, the antidot has a single ground state. A similar situation was assumed in the model of Ref.~\cite{Averin_2007}. 
As we increase $L_{AD}$, the resulting entropy change will gradually change. First, we will get contributions in the $N=0$ state from both topological sectors $I,\psi$, and their relative contribution will be determined by the ratio of scaling dimensions of $h_I=0$ and $h_\psi=1/2$. In the other extreme limit where $k_B T \gg \frac{2\pi}{L_{AD}}$, the calculation proceeds in a very similar way and we can treat the AD partition functions using Eq.~(\ref{eq_chi_a}), giving ($ k_B T \ll \frac{2\pi}{L},\frac{2\pi}{L_{AD}}, E_c$)
\bea
Z_{0 \to {\scriptscriptstyle \frac{1}{4}}}(n_g)&\propto & e^{-\beta E_c (0-n_g)^2} (d_I d_I+d_\psi d_\psi) \nonumber \\
&+&
e^{-\beta E_c (\frac{1}{4}-n_g)^2} d_\sigma d_\sigma.
\eea
Interestingly, in this case, the degeneracy factor is the same for $N=0,1/4$ since $d_I d_I+d_\psi d_\psi=d_\sigma d_\sigma=2$.

\bibliography{bibliography}

\end{document}